\documentclass[conference]{IEEEtran}
\IEEEoverridecommandlockouts

\usepackage[T1]{fontenc}
\usepackage{amsmath,amssymb,amsfonts,amsthm}
\usepackage{cite}
\usepackage{graphicx}
\graphicspath{{figures/}}
\usepackage{booktabs}
\usepackage{url}
\usepackage{dblfloatfix}

\begin{document}

\title{Coordinating Distributed AI Servers with Industrial Loads to Reduce Grid-Interconnection Needs\thanks{This work was partly supported by the National Science Foundation under award ECCS-2239046 and by Columbia School of Engineering and Applied Science Interdisciplinary Research Seed Funding.}}

\author{\IEEEauthorblockN{Ruike Lyu and Bolun Xu}
\IEEEauthorblockA{Columbia University, Earth and Environmental Engineering, New York, NY, USA}
\IEEEauthorblockA{Email:  rl3763@columbia.edu, bx2177@columbia.edu}}

\maketitle

\begin{abstract}
We show that co-locating distributed AI servers with industrial demand and coordinating AI workloads with production loads can substantially reduce the additional grid interconnection that AI computing requires. Grid interconnection is increasingly constraining the expansion of large AI data centers. Factories already have grid connections sized for their peak production loads, and we propose hosting AI servers during production valleys. We formulate a joint optimization that sizes CPU and GPU servers, schedules deadline-constrained tasks, and routes tasks among factories, and use it to compare three factory-site architectures with new large data centers serving the same demand. Applying this framework to 31 Chinese manufacturing industries, we find that AI demand and production loads often do not peak together. As a result, AI servers placed at factories and scheduled according to production peaks need about 60\% less additional interconnection capacity than new large data centers, and 98\% less when tasks are routed among a group's factories.
\end{abstract}

\begin{IEEEkeywords}
AI computing load, data center flexibility, grid interconnection, industrial load, workload scheduling
\end{IEEEkeywords}

\section{Introduction}

Rapid growth in AI computing load is driving up global data-center electricity use and interconnection needs~\cite{IEA2025EnergyAI}. Because new generation and grid infrastructure take years to plan, permit, and build, large data centers can face lengthy interconnection delays~\cite{FrickSrinivasan2025LargeLoadReview}.
Efforts to ease this bottleneck focus on expanding the electricity infrastructure available to data centers and making data-center demand more flexible~\cite{ChenZheng2026DeferOrShift}. Shifting data-center workloads across time and locations has long been studied as a way to reduce electricity costs, emissions, and peak demand charges~\cite{ColangeloEtAl2026GridInteractiveAI}. More recently, such flexibility has been examined as a means of accelerating large-load interconnection and reducing the grid capacity that new data centers require~\cite{GM_Norris2025,Butt2026ICPAI}. These measures make it easier to integrate large data centers into the grid, but the AI computing load is still located at new sites, each of which needs its own grid connection.

With the growing capability and hardware efficiency of open-weight models~\cite{yang2025qwen3, DeepSeekV32_2025}, deploying local AI servers at factory sites offers a promising alternative to meet part of enterprises' own AI needs. Factory grid connections are sized for peak production loads, which provides considerable interconnection headroom during non-peak use times~\cite{lyu2026industrial}. AI servers placed at factories can use this spare capacity depending on the type of AI services and when these tasks run. Interactive services such as office assistants must respond immediately, but tasks such as R\&D simulation and production scheduling can often wait for hours and run when production loads are low. Factory-site deployment therefore has the potential to reduce the additional interconnection capacity needed for AI loads. 

This study quantifies the additional grid-interconnection capacity required by AI computing loads when they are deployed at factory sites and coordinated with production loads. It addresses a gap in the literature where commercial investment has concentrated on hyperscalers and ever-larger AI models, while the potential of efficient open-weight models running on local AI servers to meet industrial AI demand more cost-effectively and sustainably has been largely overlooked. The contributions of this paper are as follows.
\begin{itemize}
\item We formulate a sizing-and-scheduling model to jointly determine server size and AI task schedules around production loads, and use it to study various AI deployment architectures. 
\item Using projected 2030 AI demand and measured hourly production loads for 31 Chinese manufacturing industries~\cite{GM_EWELD2023}, we compare three factory-site architectures and benchmark them with centralized data center solutions. 
\item Our study shows that new large data centers would require 1,488~MW of additional interconnection capacity to serve the studied industrial AI demand, whereas factory-site deployment would require 556--576~MW, or as little as 30~MW with routing among factories. Most individual factory sites require less than 1~MW.
\end{itemize}
We formulate the problem in Section~II, and present the case study in Section~III. Section~IV concludes the paper.

\section{Problem and Model}\label{sec_model}

\subsection{Service Scope and Deployment Architectures}

We consider scenarios in which manufacturers use AI services for their own operations, which can run either on-site or in a shared computing pool. These AI services are solely for manufacturer business support. We classify these AI demands into six categories: office knowledge services, business-process agents, visual quality review, predictive maintenance and diagnosis, production scheduling and optimization, and R\&D simulations \& digital twins. Large language models and other foundation models are integral to many of these services, typically interpreting requests and orchestrating workflows. Specialized models, optimizers, and simulators perform the majority of underlying CPU and GPU computations. Demand for each class is categorized as either firm demand, which must be served immediately upon arrival, or flexible demand, which can be delayed until an intraday or multi-day (batch) deadline. Service is quantified in reference workload units, and server throughput in such units per server-hour.

\begin{figure}[!t]
\centering
\includegraphics[width=\columnwidth]{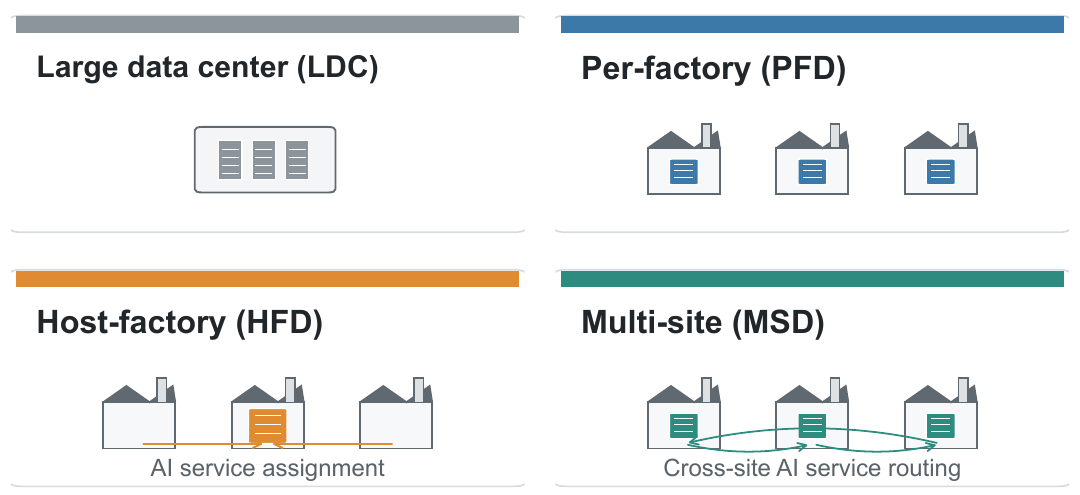}
\caption{Illustration of the four AI server deployment architectures.}
\label{fig_architectures}
\end{figure}

We consider four AI server deployment architectures (Fig.~\ref{fig_architectures}). Per-factory deployment (PFD) runs each service locally at its source factory. Host-factory deployment (HFD) runs all services for a corporate group at a single designated factory site. Multi-site deployment (MSD) allows tasks to be routed among several factories in the group within their deadlines. Large-data-center deployment (LDC) serves the same AI demand in new facilities without existing load or grid connections. Both HFD and LDC consolidate computing at a single site; however, HFD utilizes the existing grid connection of a host factory, where production load leaves part of the connection capacity unused outside peak periods. In contrast, LDC requires a new connection sized for its full AI load peak. Each factory that hosts servers is defined as a computing node, its AI load sharing the grid connection with the production load.

\subsection{Server Sizing and Scheduling Model}

\begin{figure*}[!t]
\centering
\includegraphics[width=0.75\textwidth]{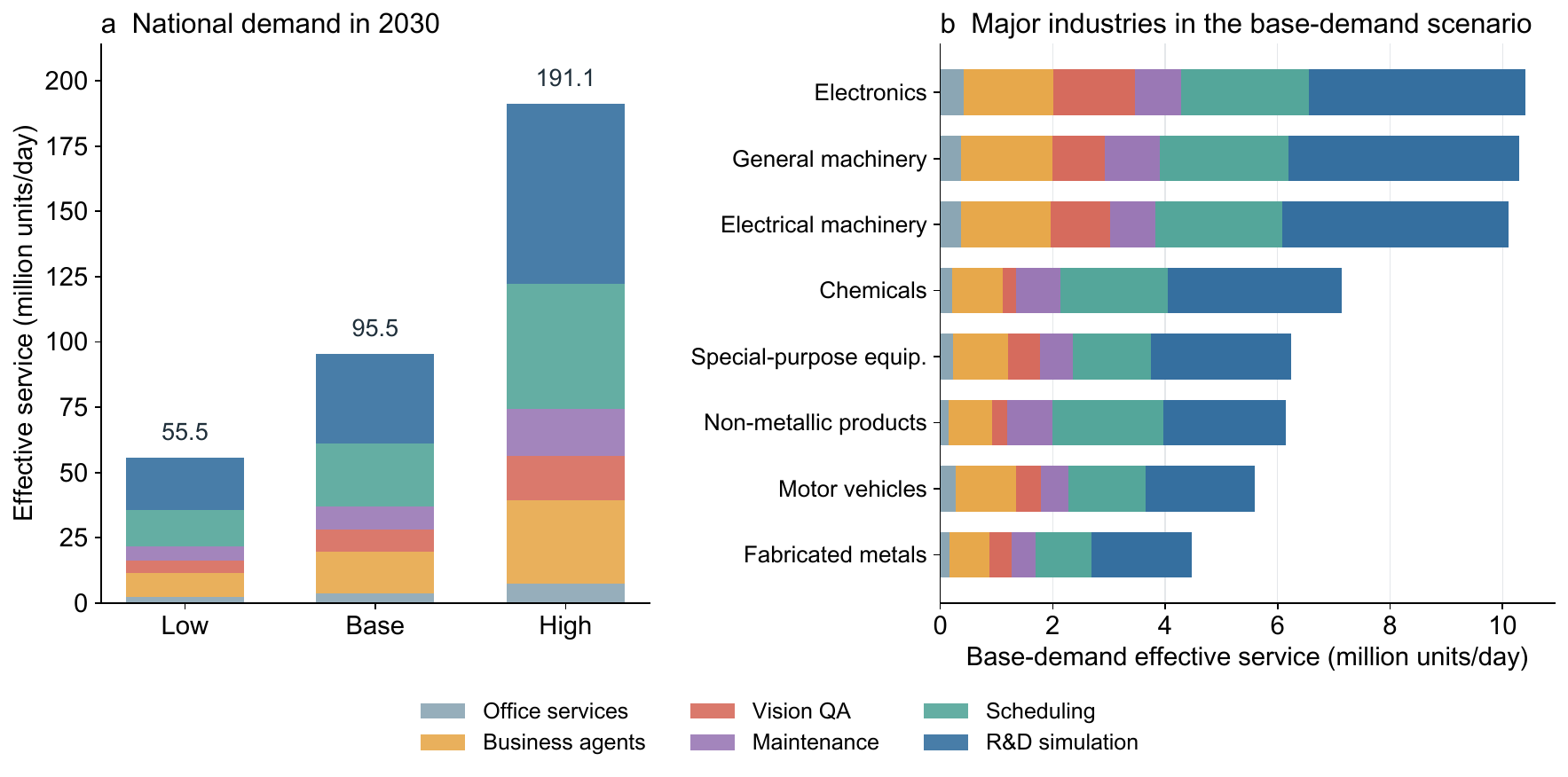}
\caption{Manufacturing AI service demand in China in 2030. (a) National demand by task class in the low-, base-, and high-demand scenarios; all results presented use the base (core) scenario. For scale, a top-down estimate derived from national projections of computing electricity use is about 14~TWh/yr~\cite{GM_CAICT2025}. (b) The eight largest industries in the base-demand scenario.}
\label{fig_demand}
\end{figure*}

For each architecture $a$ and industry, the enterprise determines the sizing of CPU and GPU servers at each node and schedules AI tasks over a representative horizon $\mathcal{T}$ to minimize annualized cost. In the initial quantification, the model is deterministic and assumes production loads $L_{n,t}$ are fixed. Therefore, any cost reduction results solely from the placement and scheduling of the AI load, rather than from industrial demand flexibility~\cite{GM_Lyu2025}. The nomenclature and problem formulation are listed below.

\noindent\textit{Indices and sets:}
\begin{IEEEdescription}[\IEEEusemathlabelsep\IEEEsetlabelwidth{$y_{k,n,t}$, $x_{j,n,t}$}]
\item[$a$] Deployment architecture (PFD, HFD, MSD, or LDC).
\item[$\mathcal{N}_a$, $n$] Computing nodes under $a$; node.
\item[$h$] Hardware type, CPU or GPU.
\item[$k$] Task class.
\item[$j$] Flexible job, with task class $k_j$.
\item[$\mathcal{T}$, $t$] Hours of the representative horizon; hour.
\item[$\mathcal{T}_j$] Hours in which job $j$ can be executed.
\end{IEEEdescription}
\noindent\textit{Decision variables:}
\begin{IEEEdescription}[\IEEEusemathlabelsep\IEEEsetlabelwidth{$y_{k,n,t}$, $x_{j,n,t}$}]
\item[$z_{n,h}$] Installed server groups (integer).
\item[$v_{n,h,t}$] Online server groups.
\item[$u_{n,h,t}$] Compute demand.
\item[$y_{k,n,t}$, $x_{j,n,t}$] Executed firm and flexible service.
\item[$p_{n,t}$] AI load (facility power).
\item[$\delta_n$] Incremental connection capacity.
\end{IEEEdescription}

Our optimization problem is formulated as follows
\begin{subequations}\label{eq_model}
\begin{align}
\min\;& \textstyle\sum_{n,h}C_hz_{n,h}+C_{\mathrm{d}}\sum_n\delta_n+Y\sum_{n,t}\Pi_tp_{n,t} \label{eq_objective}\\
\text{s.t.}\;& \textstyle\sum_{n\in\mathcal{N}_a}y_{k,n,t}=R_{k,t}, \label{eq_firm}\\
& \textstyle\sum_{n\in\mathcal{N}_a}\sum_{t\in\mathcal{T}_j}x_{j,n,t}=D_j, \label{eq_flex}\\
& y_{k,n,t}=\Omega_nR_{k,t},\ \textstyle\sum_{t\in\mathcal{T}_j}x_{j,n,t}=\Omega_nD_j, \label{eq_fixed_share}\\
& u_{n,h,t}=\textstyle\sum_kW_kQ_{k,h}\big(y_{k,n,t}+\sum_{j:k_j=k}x_{j,n,t}\big), \label{eq_compute}\\
& u_{n,h,t}\le G_hv_{n,h,t},\quad 0\le v_{n,h,t}\le z_{n,h}\in\mathbb{Z}_+, \label{eq_throughput}\\
& p_{n,t}=M_a\textstyle\sum_h\big[B_hz_{n,h}+(I_h-B_h)v_{n,h,t}\nonumber\\
&\qquad\qquad+(P_h-I_h)u_{n,h,t}/G_h\big], \label{eq_power}\\
& L_{n,t}+p_{n,t}\le K_n+\delta_n,\quad \delta_n\ge0, \label{eq_balance}\\
& z_{n,h}\ge(1+r)\textstyle\sum_{t}u_{n,h,t}/(|\mathcal{T}|G_h), \label{eq_margin}
\end{align}
\end{subequations}
where each constraint holds for all indices that appear in it.

\textit{Objective:} Equation~\eqref{eq_objective} defines the objective as the sum of the annualized cost $C_h$ for each installed server group, the maximum-demand charge $C_{\mathrm{d}}$ for incremental connection capacity, and the energy cost of the AI load at hourly prices $\Pi_t$, with the planning horizon repeated $Y=365/7$ times per year.

\textit{Service:} Constraint~\eqref{eq_firm} requires that all firm service $R_{k,t}$ of task class $k$ be executed in its arrival hour, with $y_{k,n,t}$ denoting the portion executed at node $n$. Flexible demand sharing the same task class, release hour $A_j$, and maximum delay $F_j$ is aggregated into a flexible job $j$. Constraint~\eqref{eq_flex} ensures that each job's volume $D_j$ is completed within $\mathcal{T}_j=\{(A_j+\tau)\bmod|\mathcal{T}|:\tau=0,\ldots,F_j\}$, where $x_{j,n,t}$ represents the service of job $j$ executed at node $n$ in hour $t$, and $x_{j,n,t}=0$ for $t\notin\mathcal{T}_j$.

\textit{Spatial allocation:} In per-factory and host-factory deployments, the execution location of each service is predetermined, resulting in each node executing a fixed share $\Omega_n$ of every firm task-hour and every flexible job~\eqref{eq_fixed_share}. In contrast, multi-site deployment removes this constraint, allowing the time and location of each task to be determined jointly.

\textit{Server sizing:} The executed service generates compute demand $u_{n,h,t}$ on hardware $h\in\{\mathrm{cpu},\mathrm{gpu}\}$~\eqref{eq_compute}, where $W_k$ denotes the compute required per unit of service and $Q_{k,h}$ the compute routed to $h$ per unit. Each online server group $v_{n,h,t}$ can serve up to $G_h$ per hour and cannot exceed the number of installed server groups $z_{n,h}$~\eqref{eq_throughput}, which constitutes the sizing decision. Constraint~\eqref{eq_margin} sizes installed capacity at least a planning margin $r$ above average compute demand.

\textit{AI load and connection capacity:} Each server group consumes standby power $B_h$ when installed and $I_h$ when online but idle, with power consumption increasing linearly to $P_h$ at full load. The facility-power multiplier $M_a$ accounts for cooling and power delivery~\eqref{eq_power}. Constraint~\eqref{eq_balance} ensures that grid demand, defined as the sum of production load $L_{n,t}$ and AI load, remains within the existing connection capacity $K_n$ plus incremental capacity $\delta_n$. Here, $K_n=\max_tL_{n,t}$ is the factory's peak demand without AI servers, which is conservative since actual connections often have spare contract or network capacity. Given $C_{\mathrm{d}}>0$, the optimal $\delta_n=\max\{0,\max_t(L_{n,t}+p_{n,t})-K_n\}$ is the increase in peak demand due to AI, and the national requirement is $\sum_n\delta_n$. For new large data centers, $L_{n,t}=0$ and $K_n=0$, so the entire AI peak is additional. The model can therefore avoid $\delta_n$ by shifting or routing work away from production peaks, at the cost of more servers or higher-priced energy hours.

\begin{table}[!t]
\caption{Deployment Architectures as Model Instances}
\label{tab_architectures}
\centering
\footnotesize
\setlength{\tabcolsep}{4pt}
\begin{tabular}{lllll}
\toprule
 & PFD & HFD & MSD & LDC\\
\midrule
Nodes $\mathcal{N}_a$ & each factory & one host & $\le5$ hosts & one pool\\
Production load $L_{n,t}$ & own site & host site & host sites & 0\\
Routing & fixed, \eqref{eq_fixed_share} & none & optimized & none\\
Facility multiplier $M_a$ & 1.30 & 1.30 & 1.30 & 1.22\\
\bottomrule
\end{tabular}
\end{table}

Table~\ref{tab_architectures} provides a summary of the four instances. In the large-data-center deployment, the demand from all 31 industries is aggregated at a single node. This pooling is based on average demand and is therefore conservative.

\section{Case Study}\label{sec_case}

\begin{table}[!t]
\caption{Additional Grid-Interconnection Capacity and Battery Alternative Under the Core Scenario}
\label{tab_totals}
\centering
\footnotesize
\setlength{\tabcolsep}{4pt}
\begin{tabular}{lrrrr}
\toprule
 & Capacity & Reduction & Battery & Battery cost\\
 & (MW) & vs.\ LDC & (MWh) & (CNY bn)\\
\midrule
LDC & 1,488 & -- & -- & --\\
PFD & 556 & 63\% & 1,080 & 0.71--1.09\\
HFD & 576 & 61\% & 1,583 & 0.93--1.49\\
MSD & 30 & 98\% & 46 & 0.03--0.05\\
\bottomrule
\multicolumn{5}{l}{\scriptsize Battery power equals the additional capacity.}
\end{tabular}
\end{table}

\begin{figure}[!t]
\centering
\includegraphics[width=0.75\columnwidth]{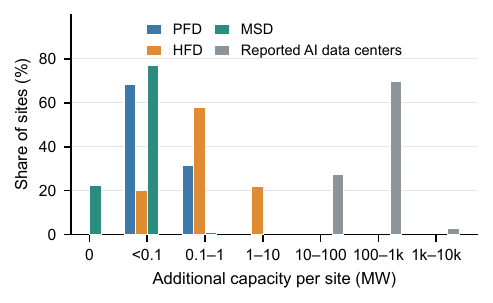}
\caption{Distribution of factory sites by additional capacity range under the core scenario, in comparison with the peak facility power of 69 publicly reported AI data centers and campuses (gray; Epoch AI, 1.2$\times$ IT power).}
\label{fig_sites}
\end{figure}

\begin{figure*}[!t]
\centering
\includegraphics[width=0.94\textwidth]{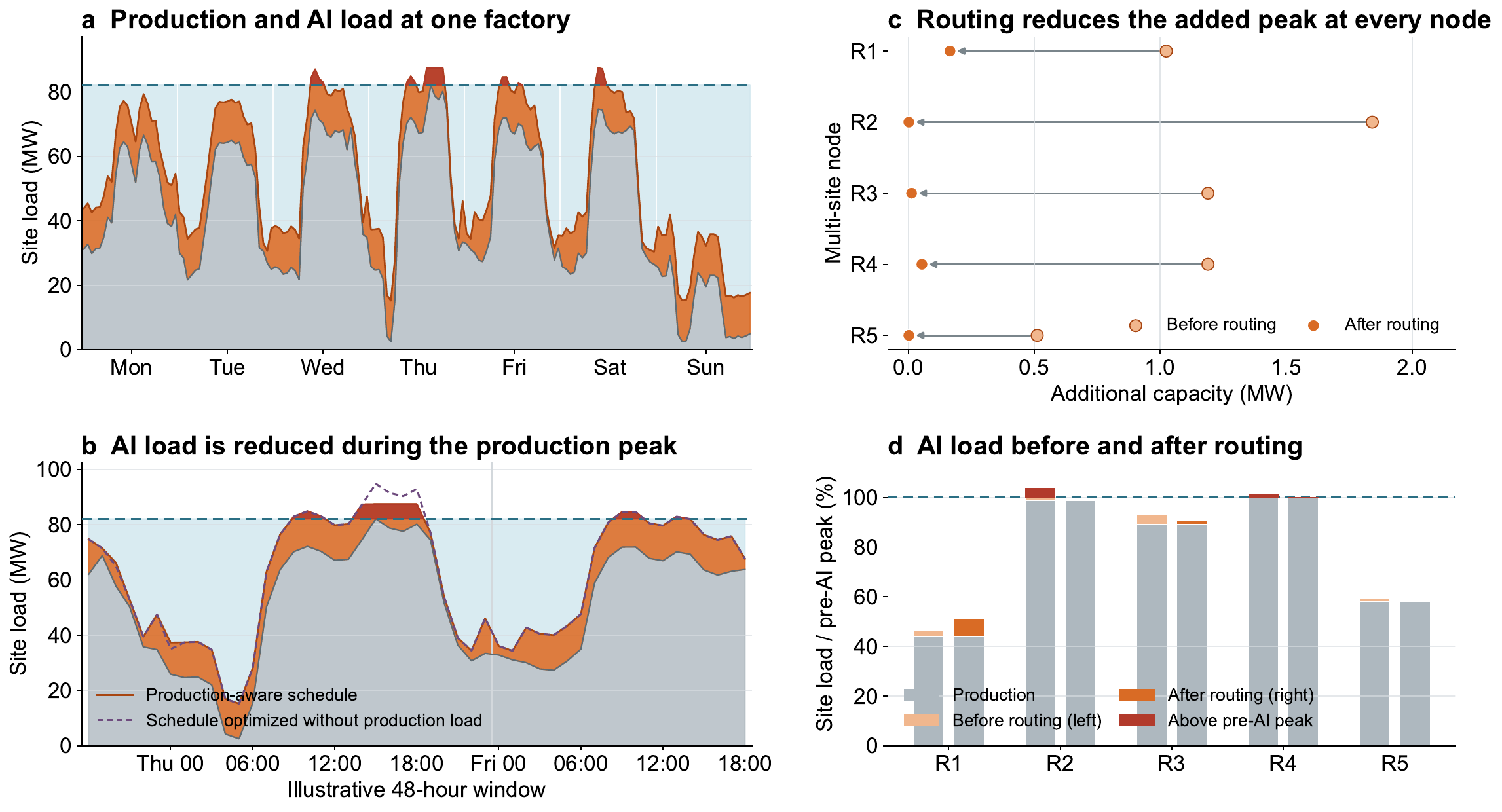}
\caption{Non-coincident peaks within and across factories in the vehicle-manufacturing group. (a) Production load (gray) and AI load (orange) at factory R4 during a representative week under host-factory deployment, in which R4 hosts the AI demand of all nine production sites in the group. The dashed line marks the pre-AI production peak, pale blue denotes the remaining headroom, and red highlights load exceeding that peak. (b) A 48-hour interval surrounding the weekly production peak; the purple dashed line represents the combined load when the AI schedule is optimized independently of production load. (c) Additional capacity at each of the five multi-site nodes before and after cross-site routing. (d) Production and AI load at each node before (left) and after (right) routing, normalized by the node's pre-AI peak.}
\label{fig_load_alignment}
\end{figure*}

We consider a 2030 AI demand scenario estimated bottom-up for 31 Chinese manufacturing industries, based on enterprise adoption, task coverage, and service intensity, consistent with a reported national adoption rate above 30\%~\cite{GM_MIIT2026}. In the core scenario, demand reaches 96 million service units per day (Fig.~\ref{fig_demand}), or approximately 12~TWh/yr of AI facility electricity in the model, representing about 0.25\% of the projected 2030 electricity consumption of the 31 industries~\cite{LyuXu2026DistributedAIRepo}. Due to the unavailability of enterprise logs, flexibility shares are assigned to each task class according to the nature of its work. Approximately one third of service units are firm, while the remainder can be delayed by up to 4--12~h (intraday) or 72~h (batch). R\&D simulation and production scheduling, which together account for 61\% of demand, are the most flexible, with firm shares of 10\% and 25\%, respectively. In contrast, interactive office services and visual review are 70--75\% firm. Weekly production-load shapes are derived from measured hourly profiles in EWELD~\cite{GM_EWELD2023}. The representative horizon consists of a 168-hour week of weekday-aligned profiles.

The reference GPU server group consists of a 2U server equipped with two NVIDIA L20 accelerators, drawing 1.50~kW at full load and 0.36~kW when idle. Compute per task is reduced threefold across all architectures to reflect anticipated hardware efficiency gains by 2030. Installed capacity incorporates a 15\% planning margin over average compute demand. Electricity prices are based on the Guangdong two-part tariff for industrial users~\cite{GM_GuangdongTariff2026}, including a maximum-demand charge of 36.1~CNY/(kW$\cdot$month) and hourly day-ahead energy prices. All instances are solved using Gurobi; code and data are publicly available.\footnote{\url{https://github.com/ruikelyu/distributed_LLM}}

\section{Results}\label{sec_results}

\subsection{Grid-Interconnection Requirements}

All three factory-site architectures require substantially less additional grid-interconnection capacity compared to new large data centers serving equivalent AI demand (Table~\ref{tab_totals}). National totals are 556~MW for per-factory deployment and 576~MW for host-factory deployment, representing reductions of 63\% and 61\%, respectively, relative to the 1,488~MW required by new large data centers. Multi-site deployment further reduces the requirement to 30~MW, which is 98\% lower than that of new large data centers.

Factory-site deployment also requires much less capacity at each connection point. In per-factory deployment, 68\% of sites require less than 0.1~MW of additional capacity, while 58\% of host-factory sites require between 0.1 and 1~MW. In multi-site deployment, more than 99\% of sites require less than 0.1~MW, and 22\% require no additional capacity (Fig.~\ref{fig_sites}). For comparison, 69 AI data centers and campuses in the Epoch AI database~\cite{GM_EpochAI2026}, primarily located in the United States, exhibit peak demands ranging from 21 to 1,135~MW, with a median of 182~MW. Consequently, a new AI data center typically requires a connection more than two orders of magnitude larger than that required by a factory site.

\subsection{Mechanisms of Capacity Reduction}\label{sec_mechanisms}

Here we examine why co-locating AI computing with production loads can reduce required network capacity.

\textit{1) Non-coincident AI and production peaks:} Factory grid connections are designed to accommodate production peaks, resulting in unused connection capacity during off-peak periods that can be utilized to host AI servers (Fig.~\ref{fig_load_alignment}a,b). In the absence of production loads at host sites, the entire AI peak constitutes an additional demand, requiring 1,585~MW for host-factory deployment, which is marginally higher than the requirement for new large data centers due to a higher facility multiplier. Co-location reduces this to 576~MW (64\%). Without task shifting ($F_j=0$), the standalone requirement is 2,372~MW, and co-location still reduces it to 1,343~MW (43\%), as AI and production peaks do not fully overlap.

\textit{2) Non-coincident peaks across factories:} Factories encounter production peaks at different times, leading to variable spare connection capacity across sites throughout the day. Routing computing tasks to sites with available capacity enables more efficient utilization of existing infrastructure (Fig.~\ref{fig_load_alignment}c,d). If tasks remain fixed to their origin nodes as defined in~\eqref{eq_fixed_share}, national multi-site deployment requires 562~MW. Permitting routing among these nodes reduces the requirement to 30~MW, representing a 95\% reduction. This analysis assumes unrestricted routing of all tasks, without network or data-residency limitations; therefore, the 95\% reduction illustrates the upper-bound benefit of cross-site routing.

\subsection{Possible Infrastructure Reductions}\label{sec_infra}

Most factory sites require less than 1~MW of additional network capacity, thereby avoiding the large-load interconnection processes encountered by new AI data centers~\cite{FrickSrinivasan2025LargeLoadReview}. These additional peaks can also be shaved with batteries that factories install and operate behind the meter~\cite{logan2026nonparametric}. We consider one battery sized at each AI host site to cover its largest daily excess above the pre-AI peak (only 1.6\% of per-factory sites, where AI load stays above the pre-AI peak for most of the week, cannot fully shave the peak increase), based on Chinese battery power and energy costs~\cite{GM_CNESA2026,LyuXu2026DistributedAIRepo}. For multi-site deployment, batteries cost CNY~0.05~billion or less (Table~\ref{tab_totals}). In comparison, the 220-kV substation and lines of a 300-MVA data center in Datong cost CNY~0.23~billion~\cite{GM_Datong2025}, or CNY~1.2~billion when scaled to 1,488~MW. 

Factory-site deployment further reduces other infrastructure requirements. Per-factory deployment uses 0.47~million m$^3$ nationally of on-site cooling and humidification water per year, about 85\% less than new large data centers providing equivalent service, because air-cooled factory server rooms are assumed to use 0.05~L/kWh of IT energy, compared to about 0.33~L/kWh reported for air-cooled large data centers~\cite{GM_AlibabaESG2025}. Per-factory deployment requires 0.92~million m$^2$ of floor area, or about 120~m$^2$ per site, which can likely be accommodated within existing industrial sites. In contrast, new large data centers require 1.34~million m$^2$ of floor area and 0.38~million m$^2$ of land conversion, estimated from public project filings~\cite{LyuXu2026DistributedAIRepo}. These reductions need not increase enterprise payments, which at public prices (Aug.~2026) are CNY~55~billion per year at factory sites and CNY~95--101~billion for equivalent cloud procurement.

\section{Conclusion}\label{sec_conclusion}

We show that locating AI servers at factory sites reduces the additional interconnection capacity needed to serve manufacturers' AI demand by 60\% compared to building new large data centers, and by up to 98\% when AI tasks can be routed among deployed servers across multiple factories under the same enterprise. This is because AI and production loads do not peak simultaneously, and flexible AI tasks can be scheduled outside production peak periods. In addition, since factories’ production loads peak at different times, AI tasks can be assigned in real time to sites with network headroom, further reducing needed interconnection capacity.

Beyond grid interconnection capacity, deploying distributed AI servers could also reduce land and water use compared with large data centers. Future research should further examine how differences in computing capability and efficiency between distributed and centralized architectures, temporal variations and uncertainties in industrial production and AI demand affect the achievable infrastructure savings and benefits of distributed AI computing.

\bibliographystyle{IEEEtran_tsg}
\bibliography{tsg_ctl,reference}

\begin{thebibliography}{10}
\providecommand{\url}[1]{#1}
\csname url@samestyle\endcsname
\providecommand{\newblock}{\relax}
\providecommand{\bibinfo}[2]{#2}
\providecommand{\BIBentrySTDinterwordspacing}{\spaceskip=0pt\relax}
\providecommand{\BIBentryALTinterwordstretchfactor}{4}
\providecommand{\BIBentryALTinterwordspacing}{\spaceskip=\fontdimen2\font plus
\BIBentryALTinterwordstretchfactor\fontdimen3\font minus
  \fontdimen4\font\relax}
\providecommand{\BIBforeignlanguage}[2]{{%
\expandafter\ifx\csname l@#1\endcsname\relax
\typeout{** WARNING: IEEEtran.bst: No hyphenation pattern has been}%
\typeout{** loaded for the language `#1'. Using the pattern for}%
\typeout{** the default language instead.}%
\else
\language=\csname l@#1\endcsname
\fi
#2}}
\providecommand{\BIBdecl}{\relax}
\BIBdecl

\bibitem{IEA2025EnergyAI}
\BIBentryALTinterwordspacing
{International Energy Agency}, ``Energy and {AI},'' International Energy
  Agency, Paris, Tech. Rep., 2025. [Online]. Available:
  \url{https://www.iea.org/reports/energy-and-ai}
\BIBentrySTDinterwordspacing

\bibitem{FrickSrinivasan2025LargeLoadReview}
\BIBentryALTinterwordspacing
N.~M. Frick and V.~Srinivasan, ``2025 large load literature review,'' Lawrence
  Berkeley National Laboratory, Tech. Rep., Dec. 2025. [Online]. Available:
  \url{https://eta-publications.lbl.gov/publications/large-load-literature-review}
\BIBentrySTDinterwordspacing

\bibitem{ChenZheng2026DeferOrShift}
Y.~Chen and X.~Zheng, ``To defer or to shift? {The} role of {AI} data center
  flexibility on grid interconnection,'' arXiv:2604.05376, 2026.

\bibitem{ColangeloEtAl2026GridInteractiveAI}
P.~Colangelo, A.~K. Coskun, J.~Megrue, C.~Roberts, S.~Sengupta, V.~Sivaram,
  E.~Tiao, A.~Vijaykar, C.~Williams, D.~C. Wilson, B.~Records, Z.~MacFarland,
  D.~Dreiling, N.~Morey, A.~Ratnayake, and B.~Vairamohan, ``{AI} data centres
  as grid-interactive assets,'' \emph{Nature Energy}, vol.~11, no.~2, pp.
  254--261, 2026.

\bibitem{GM_Norris2025}
T.~H. Norris, T.~Profeta, D.~Pati{\~n}o-Echeverri, and A.~Cowie-Haskell,
  ``Rethinking load growth: Assessing the potential for integration of large
  flexible loads in {US} power systems,'' Duke Univ., Tech. Rep., 2025.

\bibitem{Butt2026ICPAI}
H.~Z. Butt, R.~Fatima, and X.~Li, ``Minimizing grid interconnection capacity
  requirements for {AI} data centers: A developer-side planning framework with
  onsite resources and workload flexibility,'' arXiv:2608.29359, 2026.

\bibitem{yang2025qwen3}
A.~Yang, A.~Li, B.~Yang, B.~Zhang, B.~Hui, B.~Zheng, B.~Yu, C.~Gao, C.~Huang,
  C.~Lv \emph{et~al.}, ``Qwen3 technical report,'' \emph{arXiv preprint
  arXiv:2505.09388}, 2025.

\bibitem{DeepSeekV32_2025}
{DeepSeek-AI}, ``{DeepSeek-V3.2}: Pushing the frontier of open large language
  models,'' \emph{arXiv preprint arXiv:2512.02556}, 2025.

\bibitem{lyu2026industrial}
R.~Lyu, A.~Li, J.~Wang, H.~Luo, Y.~Shen, H.~Guo, E.~Du, C.~Kang, and
  J.~Jenkins, ``Industrial overcapacity can enable seasonal flexibility in
  electricity use,'' \emph{Nature Energy}, pp. 1--13, 2026.

\bibitem{GM_EWELD2023}
G.~Liu, J.~Liu, Y.~Bai, C.~Wang, H.~Wang, H.~Zhao, G.~Liang, J.~Zhao, and
  J.~Qiu, ``{EWELD}: A large-scale industrial and commercial load dataset in
  extreme weather events,'' \emph{Sci. Data}, vol.~10, p. 615, 2023.

\bibitem{GM_CAICT2025}
{China Academy of Information and Communications Technology}, ``Research report
  on coordinated development of computing power and electric power (2025),''
  CAICT, Tech. Rep., 2025, in Chinese.

\bibitem{GM_Lyu2025}
R.~Lyu, H.~Guo, G.~Strbac, and C.~Kang, ``Data-driven dimension reduction for
  industrial load modeling using inverse optimization,'' \emph{IEEE Trans.
  Smart Grid}, vol.~16, no.~3, pp. 2695--2698, 2025.

\bibitem{GM_MIIT2026}
{MIIT of China}, ``Symposium on manufacturing digital transformation
  assessment,'' Jul. 2026, in Chinese.

\bibitem{LyuXu2026DistributedAIRepo}
``{distributed\_LLM},'' \url{github.com/ruikelyu/distributed_LLM}, 2026.

\bibitem{GM_GuangdongTariff2026}
{Guangdong Power Grid Co., Ltd.}, ``Agency power-purchase prices for industrial
  and commercial users, {April} 2026,'' 2026, in Chinese.

\bibitem{GM_EpochAI2026}
{Epoch AI}, ``{AI} data centers,'' Dataset, 2026.

\bibitem{logan2026nonparametric}
\BIBentryALTinterwordspacing
E.~Logan, N.~Qi, and B.~Xu, ``Nonparametric kernel regression for coordinated
  energy storage peak shaving with stacked services,'' in \emph{2026 IEEE Power
  \& Energy Society General Meeting (PESGM)}.\hskip 1em plus 0.5em minus
  0.4em\relax IEEE, 2026.
\BIBentrySTDinterwordspacing

\bibitem{GM_CNESA2026}
{CNESA}, ``Energy storage bid prices, {Sep.} 2026,'' 2026, in Chinese.

\bibitem{GM_Datong2025}
{Datong Ecology and Environment Bureau}, ``Approval of a data-center 220-{kV}
  grid project,'' 2025, in Chinese.

\bibitem{GM_AlibabaESG2025}
{Alibaba Group}, ``{ESG} report, fiscal year 2025,'' 2025.

\end{thebibliography}

\end{document}